\documentclass[aps,prd,onecolumn,superscriptaddress,showpacs,showkeys]{revtex4}

\usepackage{slashed}
\usepackage{color}
\usepackage{graphicx}
\usepackage{amsmath}
\usepackage{epsfig}
\usepackage{subfigure}
\def\Comment#1{}
\newcommand{\bean}{\begin{eqnarray*}}
\newcommand{\eean}{\end{eqnarray*}}

\newcommand{\gapproxeq}{\lower
.7ex\hbox{$\;\stackrel{\textstyle >}{\sim}\;$}}
\newcommand{\lapproxeq}{\lower
.7ex\hbox{$\;\stackrel{\textstyle <}{\sim}\;$}}

\newcommand\lsim{\mathrel{\rlap{\lower4pt\hbox{\hskip1pt$\sim$}}
    \raise1pt\hbox{$<$}}}
\newcommand\gsim{\mathrel{\rlap{\lower4pt\hbox{\hskip1pt$\sim$}}
    \raise1pt\hbox{$>$}}}
\newcommand{\ba}{\begin{array}}
\newcommand{\ea}{\end{array}}
\newcommand{\nn}{\nonumber}

\newcommand{\be}{\begin{small}\begin{equation}}
\newcommand{\ee}{\end{equation}\end{small}}
\newcommand{\bear}{\begin{small}\begin{eqnarray}}
\newcommand{\eear}{\end{eqnarray}\end{small}}
\newcommand{\mL}{\mathcal{L}}

\newcommand{\mF}{\mathcal{F}}

\def\bat{\begin{array}{cc}}

\newcommand{\Frac}[2]{\frac{\displaystyle #1}{\displaystyle #2}}

\usepackage{color}

\newcommand{\sm}[1]{{\color{blue} #1}}

\newcommand{\bea}{\begin{small}\begin{eqnarray}}
\newcommand{\eea}{\end{eqnarray}\end{small}}

\begin{document}
\begin{flushright}
IFT-UAM/CSIC-14-115
\\
FTUAM-14-48\\
\today
\end{flushright}


\title{{ \Large\bf A Refined Analysis on the $X(3872)$ Resonance}   }


\author{Ce Meng}
\email[]{mengce75@pku.edu.cn}
\affiliation{Department of Physics, Peking University, Beijing 100871, China}

\author{Juan~Jos\'e~Sanz-Cillero}
\email[]{juanj.sanz@uam.es}
\affiliation{Departamento de F\'isica Te\'orica and Instituto de F\'isica Te\'orica, IFT-UAM/CSIC
     Universidad Aut\'onoma de Madrid, Cantoblanco, Madrid, Spain}

\author{Meng Shi}
\email[]{shimeng1031@pku.edu.cn}
\affiliation{Department of Physics and State Key Laboratory of Nuclear Physics and Technology,
Peking University, Beijing 100871, P.R.~China}

\author{De-Liang Yao}
\email[]{d.yao@fz-juelich.de}
 \affiliation{Department of Physics and State Key Laboratory of Nuclear Physics and Technology,
Peking University, Beijing 100871, P.R.~China}
 \affiliation{Institute for Advanced Simulation, Institut f{\"u}r Kernphysik and
J\"ulich Center for Hadron Physics, Forschungszentrum J{\"u}lich, D-52425 J{\"u}lich, Germany}

\author{Han-Qing~Zheng}
\email[]{zhenghq@pku.edu.cn}
\affiliation{Department of Physics and State Key Laboratory of Nuclear Physics and Technology,
Peking University, Beijing 100871, P.R.~China}
 \affiliation{Collaborative Innovation Center of Quantum Matter, Beijing, People's Republic of China}



\begin{abstract}
We study the property of the $X(3872)$ meson by analyzing the $B\to K D\bar D^*$ and
$B\to K J/\psi \pi^+\pi^-$ decay processes.
The competition between the rescattering mediated through a  Breit--Wigner resonance
and the rescattering generated from a local $D\bar{D}^* \to D\bar{D}^*$  interaction is carefully studied through an effective lagrangian approach.
Three different  fits are performed:
pure Breit-Wigner case, pure $D\bar{D}^*$ molecule  case with only local rescattering vertices (generated by the loop chain), and the mixed case.
It is found that data supports the picture where X(3872) is mainly a ($\bar cc$) Breit--Wigner resonance with a small contribution to the self--energy  generated by $\bar DD^*$ final state interaction. For our optimal fit, the pole mass and width are found to be: $M_X=3871.2\pm0.7$\mbox{MeV} and $\Gamma_X=6.5\pm1.2$\mbox{MeV}.
\end{abstract}


\maketitle

\section{Introduction}
The $X(3872)$ is a narrow resonance close to the $D^0\bar D^{*0}$ threshold,
 which was first observed in $B^{\pm}\rightarrow K^{\pm} J/\psi\pi^+\pi^-$
 by the BELLE Collaboration~\cite{BELLE1},
 and later confirmed by CDF~\cite{CDF1}, D0~\cite{D01}
 and BABAR Collaborations~\cite{BABAR1}.
 It has also been observed at LHCb~\cite{LHCb1} and CMS~\cite{CMS1}.
 The new results of Belle show a mass
 $m_{X(3872)}=3871.85 \pm 0.27(stat.)\pm 0.19(syst.)$~\mbox{MeV}
 and width less than $1.2$~\mbox{MeV}~\cite{BELLE3}.
 A recent angular distribution analysis of the $X\to J/\Psi\pi^+\pi^-$ decay
 by LHCb has determined the $X(3872)$  quantum numbers to be $J^{PC}=1^{++}$~\cite{LHCb2}.

The decay of the X(3872) including $J/\psi\pi^+\pi^-$ and $D^0\bar D^{*0}$ and  $\bar D^0D^{*0}$ final states are studied by BESIII, BABAR and BELLE~\cite{BESIII,BABAR2,BABAR3,BELLE4,BELLE2,BABAR4}.
Furthermore, other $X(3872)$ decay channels that have been observed experimentally are
$J/\psi\pi^+\pi^-\pi^0$~\cite{psigamma},
$J/\psi\gamma$~\cite{psi2sgamma} and $\psi'\gamma$~\cite{psi2sgamma},
with  relative branching ratios
\bea
\frac{Br(X\rightarrow J/\psi\pi^+\pi^-\pi^0)}{Br(X\rightarrow J/\psi\pi^+\pi^-)}&=&1.0\pm0.4\pm0.3,\label{ratio2piover3pi}\\
\frac{Br(X\rightarrow J/\psi\gamma)}{Br(X\rightarrow J/\psi\pi^+\pi^-)}&=&0.33\pm 0.12,\\
\frac{Br(X\rightarrow \psi'\gamma)}{Br(X\rightarrow J/\psi\pi^+\pi^-)}&=&1.1\pm 0.4.
\eea
The decay mode $\psi'\gamma$ was further confirmed by the LHCb Collaboration recently~\cite{LHCb:psi2sgamma}. As for the hadronic transition modes, the dipion spectrum in the $J/\Psi\pi^+\pi^-$ is mainly given by $\rho^0$ resonance
whereas  the tripion spectrum in $J/\Psi\pi^+\pi^-\pi^0$ comes mainly from
the $\omega$ meson. The ratio in Eq.~(\ref{ratio2piover3pi}) shows that these two
processes are of the  same order. One should note that the threshold of $J/\psi\omega$ is about 8 MeV higher than $m_{X(3872)}$, and the width of $\omega$ is only about 8 MeV~\cite{PDG2014}.
Thus, the isospin symmetry breaking is not as serious as that shown in Eq.~(\ref{ratio2piover3pi}) since the phase space of $J/\psi\omega$ decay mode is extremely suppressed compared with that of $J/\psi\rho^0$. Moreover, since the mass of $X(3872)$ is very close to the threshold of $D^0\bar D^{*0}$ but not to that of $D^+D^{*-}$, the rescattering effects through the $D^{(*)}\bar{D}^{(*)}$ loops can generate large isospin symmetry breaking at the amplitude level, and the number in Eq.~(\ref{ratio2piover3pi}) can be roughly accounted for even if the original decay particle has isospin $I=0$~\cite{ccbar3}.

 On the theory side, the nature of the $X(3872)$ is still a controversial issue,
 where different approaches  have not  reached yet a full agreement.
 The analysis of Ref.~\cite{boundstate1,boundstate2,boundstate3,boundstate4} favors
 a $D^0\bar D^{*0}/\bar D^0D^{*0}$ bound state,
 as the $X(3872)$ mass  is very close to the  $D^0\bar D^{*0}$ threshold.
 Other works describe the $X(3872)$ as
 a $D^0\bar D^{*0}/\bar D^0D^{*0}$ virtual state~\cite{virtualstate},
 a tetraquark~{\cite{tetraquark}} or a hybrid state~\cite{hybrid}.
 On the other hand, it has also been considered as a mixture of a charmonium $\chi_{c1}'=\chi_{c1}(2P)$ with a $D^0\bar D^{*0}/\bar D^0D^{*0}$ component~\cite{ccbar1,ccbar2}. The mixing can be induced by the coupled-channel effects, and the S-wave $\chi_{c1}'-D\bar D^{*}/\bar DD^{*}$ coupling can also explain the closeness of $m_{X(3872)}$ to the threshold of $D^0\bar D^{*0}$ naturally~\cite{LMC:coupledchannel,Simonov:coupledchannel}. Furthermore, the existence of the substantial $\chi_{c1}'$ component in the $X(3872)$ state is supported by the analyses of the lager production rates of $X(3872)$ both in $B$ decays~\cite{ccbar1,ccbar0} and at hadron colliders~\cite{hanhao}.

In Ref.~\cite{ZhangO}, it is proposed to use the pole counting rule~\cite{morgan92} to study the nature of X(3872).
A couple channel Breit--Wigner propagator is used to describe X(3872) and it is found that two nearby poles are needed in order to describe
data. Based on this it is argued that the X(3872) is mainly of $\bar cc$ nature heavily renormalized by $\bar DD^*$ loop. However, Ref.~\cite{ZhangO} did not consider the impact effect of the bubble chain generated by $\bar DD^*$ loops, which may generate a molecular type pole. Hence it might have been argued that the conclusion made in Ref.~\cite{ZhangO} was not general. The purpose of this paper is to  extend the analysis of Ref.~\cite{ZhangO}
by further including the effect of a contact $\bar DD^*$. As we will see later, the major conclusions obtained in Ref.~\cite{ZhangO} remain unchanged.

In this paper, we  only focus on $D^0\bar{D}^{*0}$ and $J/\Psi\pi^+\pi^-$
final states. An effective lagrangian is constructed to calculate the $X(3872)$ decay
into $D^0\bar{D}^{*0}$ and $J/\Psi\pi^+\pi^-$.
Three different scenarios are taken into consideration  to fit experimental data:
a single elementary particle $X(3872)$ propagating in the $s$--channel;
only $D\bar D^*$ bubble chains with contact $\bar D D^*$ rescattering;
and the mixed situation, i.e.,  an elementary $X(3872)$ particle combined with the effect of bubble chain.
In Sec.~II, the effective lagrangian is introduced and
the $B^+\rightarrow K^+ D^0\bar D^{*0}$
and $B^+\rightarrow K^+ J/\Psi\pi^+\pi^-$  amplitudes are calculated.
In Sec.~III, the numerical fits are performed and the resonance poles are analyzed.
A brief summary is provided in Sec.~IV.
Minor technical details, such as the suppression of the longitudinal component of the amplitude with respect to the transverse one near threshold, are relegated to the
appendixes.

\section{Theoretical analysis}
\label{sec.theory}

\subsection{The Effective Lagrangian}

The $X(3872)$ has been identified as a $s$--wave resonance in the
$D^0\bar D^{*0}$ and $\bar D^0 D^{*0}$ final states with isospin $I=0$, and the similar situation for the p-wave $D\bar D$ final states near the $\psi(3770)$ was studied in~\cite{GYCheng,Achasov}
Likewise, as discussed in the introduction, we will assume $X(3872)$
to be an axial-vector resonance, with  $J^{PC}=1^{++}$.
To simplify the notations, from now on the channels $D^0\bar D^{*0}/\bar D^0 D^{*0}$ are
labeled just as $D^0\bar D^{*0}$, and the channels $D^+D^{*-}/D^-D^{*+}$ are labeled
as $D^+D^{*-}$ (unless specifically stated otherwise). Likewise,
when the two channels $D^0\bar D^{*0}/\bar D^0 D^{*0}$ and $D^+D^{*-}/D^-D^{*+}$
appear together, they are labeled as $D\bar D^*$ in what follows.

In this section we construct the effective lagrangian of the interactions between
X(3872) and other particles. The lagrangian of $D\bar D^{*}$ interactions has been
constructed by ~\cite{Lag1, Lag2, Fleming}.
However,  operators such as $BKX$, $BKD\bar D^{*}$, are not considered previously and
only constant form factors were used in these previous calculation missing information from $B$ decay vertex.

We will consider a model  written in a relativistic form but intended
 for the description of $D\bar{D}^*$   and $J/\psi \pi^+\pi^-$ invariant energies close
 to the $D^0\bar{D}^{*\, 0}$  production threshold.
 We begin by constructing operators in our lagrangian with the lowest number of derivatives
 and fulfilling invariance under $C$, $P$ and isospin symmetry.
 Hence, the interaction between the $X(3872)$ and the $D\bar{D}^*$ pair
 will occur through the combination of isospin, $C$ and $P$,
\bea
X\quad \sim \quad
 \frac{1}{\sqrt{2}}\, ( \, \bar D D^{*} \, - \,    \bar D^{*} D\, )  \, ,
\label{eq.X-DD}
\eea
where the minus sign stems from the positive $X(3872)$ $C$--parity and the usual assignment
for $D\sim i\bar{q}\gamma_5 c$ and $D^{*\mu}\sim \bar{q}\gamma^\mu c$,
with $C D^j C^{-1}= D^{j \, \dagger} $ and $CD^{j\, *}_\mu C^{-1} = - D_\mu^{j\, *\,\dagger}$,     
where $j$ indicates the type of $D$ or $D^*$ meson (e.g. $D^j=D^0,D^+$,etc)~\cite{Bratten}.
The $SU(2)$ vectors $D$ and $D^*$ gather the isospin doublets~\cite{Doublet},
${D=\left(  - D^+  \atop D^0\right)}$, $D^*=\left(  -D^{*+}   \atop D^{*0}\right)$,
with the transposed conjugates
$\bar D=(-D^-\quad \bar D^0)$, $\bar D^*=(-D^{*-}\quad\bar D^{*0})$.
 Hence, following the previous symmetry prescriptions
 we consider the isospin, $C$ and $P$ invariant effective lagrangian given by the operators,
  \bea\label{lagrangian}
 \mathcal{L}_{D\bar D^*}&=&\lambda_1(\bar D^{*\mu}D\bar D^{*\mu}D+\bar D D^{*\mu}\bar D D^{*\mu})
+\lambda_2(\bar D^{*\mu}D\bar D D^{*\mu}) \, ,\nn\\
\mathcal{L}_{XD\bar D^*}&=&g_1X^{\mu}(\bar{D}D^*_{\mu}-\bar{D}^*_{\mu}D),\nn\\
\mathcal{L}_{B X K}&=&ig_2 X^{\mu} \,
 ( \overline{B} \partial_{\mu}K \, +\, \mbox{h.c.})  \, ,
\nn\\
 \mathcal{L}_{BKD\bar D^*}&=&ig_3  (\bar D D^*_{\mu}-\bar  D^*_{\mu}D)\,
   (\overline{B} \partial^{\mu}K \, +\, \mbox{h.c.}) \, ,
\eea
with the isospin doublets  $B=\left(  B^+   \atop B^0\right)$ and $K=\left( K^+   \atop K^0\right) $.
In the present model they are combined in such
a way that the charge of the outgoing kaon always coincides with the charge
of the incoming  $B$-meson, as we are interested in processes where the remaining
decay product state is neutral and isoscalar (the quantum numbers of the $X(3872)$).
In addition to~Eq.~(\ref{lagrangian}),
for the decay into $B^+\to K^+ J/\psi\rho\,(\omega)$
we have the following operators,
\bea\label{lagrangian2}
  \mathcal{L}_{X \Psi V }&=&ig_4 X^\mu\Psi^{\nu}\partial^\alpha V^\beta
  \epsilon_{\mu\nu\alpha\beta}  \, ,\nn\\
 \mathcal{L}_{\Psi V D\bar D^*}&=&ig_5 (\bar D D^{*\mu}-\bar D^{*\mu}D)
 \Psi^{\nu}\partial^\alpha V^\beta\epsilon_{\mu\nu\alpha\beta}\, ,
 \eea
 with $V$ denoting $\rho(770)$ or $\omega(782)$.
In principle, one may also have a direct $\bar BK \Psi V$ decay through the corresponding
operator. However, since we are interested in the $X(3872)$--resonant structure,
we will not discuss this term in the lagrangian since it only contributes to the background term.
The couplings have dimensions $[g_1]=E^1$ and $[g_3]=[g_5]=E^{-1}$ while the other
coupling constants  are dimensionless.

The general structure of $\mathcal{L}_{D\bar D^*}$ contains two coupling constants
$\lambda_1$ and $\lambda_2$,
being consistent with heavy quark symmetry~\cite{Lag1}.
They provide the contact $D\bar{D}^*$ rescattering.
Operators with higher derivatives are regarded as corrections in this model
and will be neglected.
It is convenient to expand the operator $\mL_{XD\bar{D}^*}$
together with the Lagrangian $\mathcal{L}_{D\bar D^*}$ in the explicit form
\bea
&& T_{D\bar D^*} =
 \left (\begin{array}{llll}
 T_{D^{*-}D^+D^{*-}D^+}& T_{D^{*-}D^+D^{*+}D^-}& T_{D^{*-}D^+\bar D^{*0}D^0}& T_{D^{*-}D^+D^{*0}\bar
 D^0}\\
 T_{D^{*+}D^-D^{*-}D^+}& T_{D^{*+}D^-D^{*+}D^-}& T_{D^{*+}D^-+\bar D^{*0}D^0}& T_{D^{*+}D^-D^{*0}\bar
 D^0}\\
 T_{\bar D^{*0}D^0D^{*-}D^+}& T_{\bar D^{*0}D^0D^{*+}D^-} &T_{\bar D^{*0}D^0\bar D^{*0}D^0}& T_{\bar D^{*0}D^0D^{*0}\bar
 D^0}\\
 T_{D^{*0}\bar D^0D^{*-}D^+}& T_{D^{*0}\bar D^0D^{*+}D^-}& T_{D^{*0}\bar D^0\bar D^{*0}D^0}& T_{D^{*0}\bar D^0D^{*0}\bar D^0}
  \end{array}\right )
 = \left(\begin{array}{llll} 2\lambda_1 & \lambda_2 & 2\lambda_1& \lambda_2 \\
\lambda_2 & 2\lambda_1 & \lambda_2& 2\lambda_1\\
 2\lambda_1 & \lambda_2 & 2\lambda_1& \lambda_2 \\
 \lambda_2 & 2\lambda_1 & \lambda_2& 2\lambda_1
  \end{array}\right) \, ,
 \nn\\ \nn\\
 && \vec{\mF}_{X^\mu \to D\bar{D}^*} = \left(\ba{c}\mF_{X^\mu\to  D^{*-} D^+} \\
 \mF_{X^\mu \to D^{*+} D^-}  \\ \mF_{X^\mu \to \bar{D}^{*0} D^0}  \\
 \mF_{X^\mu\to  D^{*0} \bar{D}^0 }   \ea \right)
 \,\,\, =\,\,\, g_1\, \vec{u}_X \, , \qquad
 \mbox{with }  \vec{u}_X=\left(\ba{c} 1 \\ -1 \\ 1\\ -1 \ea \right)\, .
 \eea
 However, in order to match  the $D^0\bar{D}^{*0}/\bar{D}^0D^{*0}$
non-relativistic effective field theory  near threshold (the minimal charm meson model)~\cite{Lag2},
one needs $\lambda_2=-2\lambda_1$. 
Hence, under this condition the $T_{D\bar D^*}$ gets the simplified form:
 \bea
 T_{D\bar D^*}\, =\, \lambda_2  \, \vec{u}_X \, \vec{u}_X^{\, T}\, .
 \eea
Notice that this contact scattering matrix $T_{D\bar{D}^*}$ projects into the flavor structure
of the $X\to D\bar{D}^*$ transition. Hence, it accounts only for the $D\bar{D}^*$
local rescattering with the quantum number of the $X(3872)$.  From now on, we
will use $\lambda_2= -2 \lambda_1$ all along the article.

 \begin{figure}[h!]
\centering{
\includegraphics[width=8cm,height=7cm]{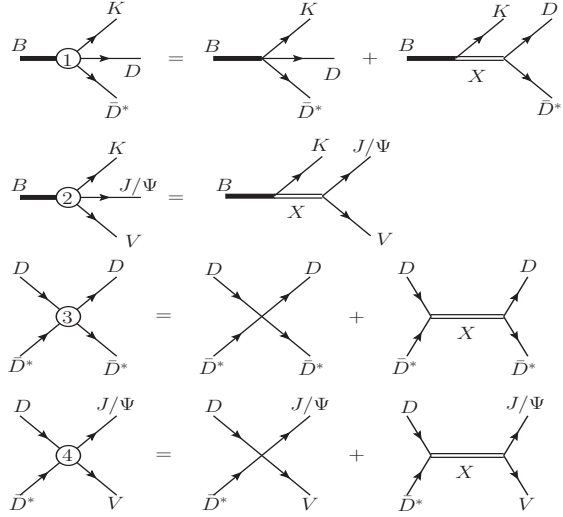}}
\caption{\label{fig1} {\small
Interaction vertices.
The blob (1) depends on the couplings $g_1\cdot g_2$ and $g_3$,
the blob (2) depends on $g_2\cdot g_4$,
the blob (3) depends on $\lambda_2$ and $g_1^2$,
and the blob (4) depends on $g_5$ and $g_1\cdot g_4$.}}
\end{figure}

\begin{figure}[h!]
\centering{
\includegraphics[width=0.65\textwidth]{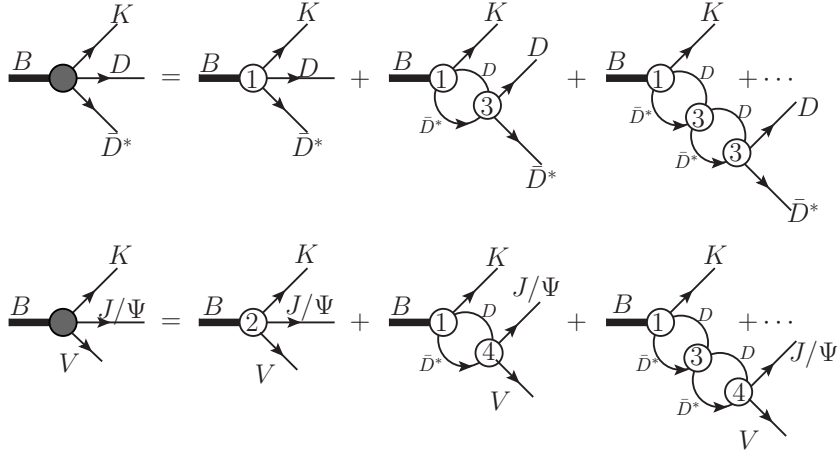}}
\caption{\label{fig2} {\small
Decay diagrams.
}}
\end{figure}

\subsection{Amplitude of $B^+\rightarrow K^+D^0\bar D^{*0}$}

Based on the lagrangian in Eqs.~(\ref{lagrangian}) and (\ref{lagrangian2}),
we extract the decay amplitude   $B^+\to K^+D^0\bar D^{*0}$.
Fig.~\ref{fig1} shows the interaction vertices and Fig.~\ref{fig2}
describes Feynmann diagrams for  $B^+\rightarrow D^0\bar D^{*0}K^+$ and
$B^+\rightarrow J/\Psi\pi^+\pi^-K^+$.
In the first line of the Fig.~\ref{fig2},  one can observe the  $D \bar{D}^{*}$
final state interaction coming  in part  from  a  bubble chain of local scatterings through the
$\lambda_2$ operator.  In general, every  rescattering is produced by two kinds of interactions:
contact interaction and $X(3872)$ exchanges in the $s$--channel.
The rescattering effective vertex denoted as (3) in Fig.~\ref{fig1} is given by
  \bea
  iA_{D\bar D^*}^{\mu\nu} &=&  i\, g^{\mu\nu}T_{D\bar D^*}\, +\, i(g_1)^2D(p^2)^{\mu\nu}
  \,   (\vec{u}_X \vec{u}_X^{\, T})\,\,\,=\,\,\,
  \bigg[ \, i\lambda_2g^{\mu\nu}+\frac{ig_1^2(g^{\mu\nu}-\frac{p^\mu p^\nu}{M_X^2})}{p^2-M_X^2}\,
  \bigg]
  \, (\vec{u}_X \vec{u}_X^{\, T})\, ,
  \eea
where $p=p_D+p_{\bar{D}^*}$  is the momentum of the $D\bar{D}^*$ system, and
$M_X$ the $X(3872)$  mass,  $\lambda_2$ provides the local scattering  of the $D\bar{D}^*$
meson pairs and $(\vec{u}_X \vec{u}_X^{\, T})$ provides the precise  structure
for the various  flavor scatterings.
For a massive particle, like the X(3872), the Proca propagator has two components,
   \bea
  D(p^2)^{\mu\nu} &=&  \frac{-i(g^{\mu\nu}-\frac{p^{\mu}p^{\nu}}{M_{X}^2})}{p^2-M_X^2}
 =\frac{-iP_{T}^{\mu\nu}(p)}{p^2-M_X^2}+\frac{iP_{L}^{\mu\nu}(p)}{M_X^2},
   \eea
with  $P_{T\mu\nu}=g_{\mu\nu}-\frac{p_\mu p_\nu}{p^2}$ and $P_{L\mu\nu}=\frac{p_\mu p_\nu}{p^2}$
the transverse and longitudinal projection operators, respectively.
The longitudinal part happens to be suppressed in the $D\bar D^*$ decay
by an extra factor $|\vec{p}_{D^*}|$ near the $D\bar D^*$ threshold and will produce a
much smaller impact. Thus, no pole will be generated in the longitudinal channel
in the neighbourhood of the $D\bar{D}^*$ threshold.
Detailed discussion on this point can be found in Appendix~\ref{appendix A}.
The effective vertices in the blobs (1) and (4) in Fig.~\ref{fig1}
also have contact interaction and $X(3872)$ exchanges.
However, in the effective vertex (2) only $X(3872)$ exchanges have been taken into account
in our model. Possible contact interactions will be treated as background to the spectrum in our later
phenomenological analysis.

After taking into account the  $D\bar{D}^*$ rescattering effect,
the $B^+\rightarrow D^0\bar{D}^{*0} K^+$  decay amplitude  can be separated
into transverse and longitudinal components,
 \bea\label{DDs}
 \mathcal{M}_{D^0\bar D^{*0}} &=&
 -\, \frac{(g_3+\frac{g_1g_2}{s-M_X^2})p_K^{\mu}\epsilon_{D^*}^{\nu}}
 {1-(i \lambda_2+i\frac{g_1^2}{s-M_X^2})\hat{\Pi}_T(s)} \, P_{T\mu\nu}(p)
 \,\,\, +\,\,\,
 \frac{(g_3+\frac{g_1g_2}{M_X^2})p_K^{\mu} \epsilon_{D^*}^{\nu}}
 {1-(i \lambda_2+i\frac{g_1^2}{M_X^2})\hat{\Pi}_L(s)} \, P_{L\mu\nu}(p) \, ,
 \eea
where $M_X$, $p_K$ and $\epsilon_{\bar D^{^*0}}$ are the mass of X(3872),
the momentum of K meson and the $\bar D^{*0}$ polarization, respectively, and $s=p^2$.
The total one-loop contributions are given by
$\hat{\Pi}_T=2(\hat{\Pi}_{T_{D^0\bar D^{*0}}}+\hat{\Pi}_{T_{D^+D^{*-}}})$ and
$\hat{\Pi}_L=2(\hat{\Pi}_{L_{D^0\bar D^{*0}}}+\hat{\Pi}_{L_{D^+D^{*-}}})$.
The factor 2 results from the identity between  the $D\bar D^{*}$ and $\bar D D^{*}$ contributions.
For instance, the $D^0\bar D^{*0}$ one loop integral is given by
 \bea
 \int\frac{d^Dk}{(2\pi)^D}
 \frac{g_{\mu\nu}-\frac{k_{\mu}k_{\nu}}{m_{D^{*0}}^2}}{(k^2-m_{D^{*0}}^2)
 ((p-k)^2-m_{D^0}^2 )}
 \,\,\, =\,\, \,P_{T\mu\nu}(p)\Pi_{T_{D^0\bar D^{*0}}}(s) \,
 +\, P_{{L\mu\nu}}(p) \Pi_{L_{D^0\bar D^{*0}}}(s)\, .
 \nn\\
 \eea

The contributions  $\Pi_{T_{D^+D^{*-}}}(s)$ and $\Pi_{L_{D^+D^{*-}}}(s)$ have similar structure
but with  charged $D\bar{D}^*$ masses instead of neutral,
having thus a different production threshold~$\sqrt{s}=3879.4$~MeV.
The  $D^0\bar D^{*0}$   threshold   is placed at $\sqrt{s}=3871.3$~MeV, 8~MeV below the charged one.
The expressions of  $\hat{\Pi}_{T_{D^0\bar D^{*0}}}(s)$ and $\hat{\Pi}_{L_{D^0\bar D^{*0}}}(s)$
are given in Appendix~\ref{appendix B}. Therein, $\hat{\Pi}_{T_{D^0\bar D^{*0}}}(s)$ and
$\hat{\Pi}_{L_{D^0\bar D^{*0}}}(s)$ are proven to be, respectively,
proportional to $-\frac{1}{16\pi}\rho(s)$ and
$\frac{1}{16\pi}\rho^3(s)$ near the $D^0\bar D^{*0}$ threshold,
with  $\rho(s)= \frac{2|\vec{p}_D|}{\sqrt{s}} =
\frac{\sqrt{(s-(m_{D^0}+m_{D^{*0}})^2)(s-(m_{D^0}-m_{D^{*0}})^2)}}{s}~$,
being
$\vec{p}_D$ here the $D$ three-momentum in the $D\bar{D}^*$ center-of-mass rest frame.

  There are also other decay channels
  with much lighter production thresholds,
  such as $J/\Psi\gamma$, etc,  which   affect the $X(3872)$  propagator.
  Since all such channel thresholds are far away from the $D^0\bar D^{*0}$ one
  and the energy region under study is a narrow range around the latter,
  the contributions to the $X(3872)$ self energies from  these channels
  can be fairly approximated as a constant.
  In order to account for these absorptive contributions,
  in the transverse part of the amplitude in Eq.\,(\ref{DDs}) we make the replacement
  \bea\label{replacement}
  s-M_X^2  \Rightarrow s-M_X^2+iM_X(\Gamma_{J/\Psi\pi\pi}(s)+\Gamma_{J/\Psi\pi\pi\pi}(s)+\Gamma_0),
  \eea
  where the effective parameters $M_X$ and $\Gamma_0$ will be  determined by our fits to
  experimental data, and $\Gamma_{J/\Psi\pi\pi}(s)$ and $\Gamma_{J/\Psi\pi\pi\pi}(s)$ are the partial widths of the X(3872) from
  $J/\Psi\rho$ and $J/\Psi\omega$ contribution. We denote the $XJ/\Psi\omega$ coupling as $g_4'$ to distinguish it from the $XJ/\Psi\rho$ coupling, denoted as $g_4$.
  The widths $\Gamma_{J/\Psi\pi\pi}(s)$ and $\Gamma_{J/\Psi\pi\pi\pi}(s)$ are
  \bea
  \Gamma_{J/\Psi\pi\pi}(s)=g_4^2\int_{2m_{\pi}}^{\sqrt{s}-m_{J/\Psi}}\frac{dm}{2\pi}
  \frac{k(m)(\frac{s\cdot k(m)^2}{m_{J/\Psi}^2}+2m_{J/\Psi}^2+2s-6\sqrt{s}k(m)+k(m)^2)\Gamma_{\rho}}{4\pi s((m-m_{\rho})^2+\Gamma_{\rho}^2/4)},
  \eea
    \bea
  \Gamma_{J/\Psi\pi\pi\pi}(s)=g_4'^2\int_{3m_{\pi}}^{\sqrt{s}-m_{J/\Psi}}\frac{dm}{2\pi}
  \frac{k(m)(\frac{s\cdot k(m)^2}{m_{J/\Psi}^2}+2m_{J/\Psi}^2+2s-6\sqrt{s}k(m)+k(m)^2)\Gamma_{\omega}}{4\pi s((m-m_{\omega})^2+\Gamma_{\omega}^2/4)},
  \eea
  where $k(m)=\sqrt{\frac{(s-(m+m_{J/\Psi})^2)(s-(m-m_{J/\Psi})^2)}{4s}}$, $m_{J/\Psi}$, $m_{\rho}$, $m_{\omega}$ and $\Gamma_{\rho}$, $\Gamma_{\omega}$ are the mass of $J/\Psi$, $\rho$, $\omega$ and the width of $\rho$,
  $\omega$ respectively.

The $D^0\bar D^{*0}$ invariant mass spectrum   is provided by
   \bea
   \frac{dN}{dm_{D^0\bar D^{*0}}}=N_1\frac{d\Gamma}{dm_{D^0D^{*0}}}=N_1\frac{2 m_{D^0\bar D^{*0}}}{(2\pi)^3
   32M_B^3}\int (|\mathcal{M}_{D^0\bar D^{*0}}|^2)dm_{DK}^2+c_1\rho_3(s),\label{invariantmassddstar}
   \eea
 where $m_{DK}$ is the invariant mass of the $D K^+$ system;
 $N_1$ is a normalization factor; $c_1$ is a constant, which multiplies the phase space
 ($\rho_3(s)=|\vec{p}_k||\vec{p}_{DD^*}|$)
 and models the background contribution.

\subsection{$B^+\rightarrow K^+J/\Psi\pi^+\pi^-$ amplitude}

 In the $J/\Psi\pi^+\pi^-K^+$ situation, we assume that the $B^+$ meson firstly decays
 into $K^+J/\Psi V$, and then $V$ decays into $\pi^+\pi^-$.
 The vertices are presented in Fig.\ref{fig1}. As explained before, no $BKJ/\Psi V$ contact interaction is considered in the present study. It is assumed to be part of the constant background term below.

 In the second line of Fig.~\ref{fig2} we show only the $D\bar{D}^*$ final state interaction.  As the energy range under study is very close to  the $D\bar D^*$ threshold,
 the $D\bar D^*$ rescattering dependence on the energy and  other channels
 are  accounted through the  constant width $\Gamma_0$ introduced in the above section in Eq.\,(\ref{replacement}) together with the $\Gamma_{J/\Psi\pi\pi}$ and $\Gamma_{J/\Psi\pi\pi\pi}$ contributions therein.

The $B^+\rightarrow J/\Psi VK^+$ amplitude is now given by the much  involved expression as the following:
  \bea \label{JPsiV}
  \mathcal{M}_{J/\Psi V}  &=&
  \nn \\
  && \hspace*{-1.4cm}
 \frac{p_K^{\mu} \epsilon_{\Psi}^\nu p_{V}^{\alpha}\epsilon_{V}^{\beta}
  \epsilon_{\rho\nu\alpha\beta}(g_2g_4(1-i\lambda_2\hat{\Pi}_T)+ig_1g_2g_5\hat{\Pi}_T+ig_3g_5(s-M_X^2)\hat{\Pi}_T+ig_1g_3g_4\hat{\Pi}_T)}
  {(s-M_X^2)(1-i\lambda_2\hat{\Pi}_T)-ig_1^2\hat{\Pi}_T}
  \, P_{T\mu}^\rho(p) \nn \\
  && \hspace*{-1.5cm}  +\, i\, g_3p_K^{\mu}\hat{\Pi}_L\epsilon_{\Psi}^\nu p_{V}^{\alpha}\epsilon_{V}^{\beta}
  \epsilon_{\rho\nu\alpha\beta}\frac{ig_5-\frac{ig_1g_4}{M_X^2}}{1-(i\lambda_2-\frac{ig_1^2}{M_X^2})\hat{\Pi}_{L}}P_{L\mu}^\rho(p)
  -\frac{ig_2p_K^{\mu}(g_4+\frac{ig_1g_5\hat{\Pi}_{L}}{1-i\lambda_2\hat{\Pi}_{L}})
  \epsilon_{\Psi}^\nu p_{V}^{\alpha}\epsilon_{V}^{\beta}
  \epsilon_{\rho\nu\alpha\beta}}{M_X^2+\frac{ig_1^2\hat{\Pi}_{L}}{1-i\lambda_2\hat{\Pi}_{L}}}
  \,P_{L\mu}^\rho(p)\, ,
  \nn\\
  \eea
where $p_V$ and $\epsilon_V$ are the momentum and the polarization of V meson,
and  $\epsilon_\Psi$ is the polarization of $J/\Psi$.
The other symbols are the same as in  Eq.~(\ref{DDs}), and the $s-M_X^2$ also needs to be replaced by $s-M_X^2+iM_X(\Gamma_{J/\Psi\pi\pi}(s)+\Gamma_{J/\Psi\pi\pi\pi}(s)+\Gamma_0)$ as before.

An adequate determination of the decay into $J/\Psi\pi^+\pi^-$  amplitude   can  be extracted from
the $B^+\rightarrow J/\Psi V K^+$ amplitude by  inserting the propagator of the $V$ particle, the $\rho(770)$,
as discussed in previous sections.
The $J/\Psi\pi^+\pi^-$ is studied for the
$B^+\rightarrow J/\Psi\pi^+\pi^-K^+$ decay.
Considering the cascade decay $B^+\rightarrow K^+ J/\Psi V\rightarrow K^+J/\Psi\pi\pi$,
the spectrum is given by~\cite{PDG2014}
  \bea\label{invariantmass:Jpipip}
  \frac{dN}{dm_{J/\Psi\pi\pi}}&=&N_2\int^{m_{J/\Psi V}-m_{J/\Psi}}_{2m_{\pi}}
  \frac{d\Gamma}{dm_{J/\Psi V}}\frac{k(m_{\pi\pi}) \,
  \Gamma_{V}}{(m_{\pi\pi}-m_{V})^2+\Gamma_{V}^2/4} dm_{\pi\pi}
  \,\,\,+\,\,\, c_2
   \nn\\
  &=&N_2\int\int\frac{2 m_{J/\Psi V}}{(2\pi)^3
   32M_B^3} |\mathcal{M}_{J/\Psi V}|^2 dm_{J/\Psi K}^2\frac{k(m_{\pi\pi})
   \Gamma_{V}}{(m_{\pi\pi}-m_{V})^2+\Gamma_{V}^2/4}dm_{\pi\pi}
   \,\,\, +\,\,\, c_2,
  \eea
  where $N_2$ is the normalization constant, $c_2$ parametrizes the background,
  $m_{\pi\pi}$ is the  $\pi^+\pi^-$ invariant mass and
  $k(m_{\pi\pi})$ is the pion three-momentum in  the $\pi^+\pi^-$ rest-frame.
  The constants $m_V$  and $\Gamma_V$ are the mass and width of the vector meson, respectively.


\section{Numerical results and pole analysis}

\subsection{Fits to the amplitudes}

In above sections we have calculated the $B^+\rightarrow D^0\bar D^{*0}K^+$
and $B^+\rightarrow J/\Psi\pi^+\pi^-$ invariant mass spectra, taking into account both the Breit--Wigner particle propagation (elementary X(3872))
and the $D\bar D^*$ bubble chain mechanisms. In this section we will carefully study the interfence and competition between the two mechanisms through a numerical fit to data. We anticipate here the $\chi^2$ prefers the elementary scenario than the molecular one, though the mixed situation (i.e., with both mechanisms involved) may not be excluded.

We perform the following three fits:
\begin{itemize}

\item{}
Case I: We assume that $D\bar D^*$ loops only renormalized the X(3872) self-energy through $XD\bar D^*$ vertex with coupling constant $g_1$. There is no
$D\bar D^*$ contact interaction and $\lambda_2=0$ in  Eqs.~(\ref{DDs})
and~(\ref{JPsiV}). This situation implies that there is a pre-existent elementary X(3872), which is not a molecular bound state generated by $D\bar D^*$ intermediate state.

\item{}
Case II:  Among the interactions in Fig.~\ref{fig1}
only the direct $D\bar D^*$ local interaction is taken into account and
intermediate $X(3872)$ exchanges  are discarded. That means setting $g_1 , \, g_2, \, g_4=0$
in amplitudes~(\ref{DDs}) and~(\ref{JPsiV}), and corresponds to the situation where the $D\bar D^*$ bubble loop chains are responsible for the experimentally observed peak, i.e., since line-shape and pole are both related but what generates both is the bubble chain.

In such a situation,
the structure of amplitudes takes then the form
\be
\Frac{f_T}{ \lambda_2^{-1} \, -\, i \hat{\Pi}_T} \,\,\,
+\,\,\, \Frac{ f_L}{\lambda_2^{-1} \, -\, i \hat{\Pi}_L } \, ,
\ee
where $f_T$ and $f_L$ denote the corresponding numerators in the amplitudes. When there is the $X$ propagation (Case I),
 the contributions from other channels, such as $J/\Psi\gamma$,
are taken into account through the constant width $\Gamma_0$ in the $X$ propagator. In the present situation (Case II) there is no intermediate $X(3872)$ elementary particle,
the contributions from other channels are taken into consideration by shifting the  coupling constant $\lambda_2$ to
 $\lambda_{eff}$:
\bea
\frac{1}{\lambda_{eff}}=\frac{1}{\lambda_2}+ic_0,
\eea
where  $c_0$ is a real constant which accounts for lower thresholds contributions. The role of $ic_0$ is to shift the pole from real axis to the complex plane (i.e., contributes a small width to the bound state).
Now the amplitude takes the form:
\bea
\Frac{f_T}{ \lambda_2^{-1} \, -\, i \hat{\Pi}_T} \,\,\,
+\,\,\, \Frac{ f_L}{\lambda_2^{-1} \, -\, i \hat{\Pi}_L } \, ,
 \rightarrow
 \frac{f_T }{\lambda_2^{-1} +ic_0- i\hat{\Pi}_T}
 +\frac{ f_L }{\lambda_2^{-1} +ic_0- i\hat{\Pi}_L}\, .
\eea
As  $\hat{\Pi}_T\propto -\rho(s)$ near the $D\bar D^*$ threshold,
the pole in the transverse component is determined by the sign of $\lambda_2$,
which will be discussed in the next subsection.

\item{}
Case III: We also tried to incorporate both fit I and fit II features by switching on all  the interactions in Figs.~\ref{fig1} and \ref{fig2}, allowing both direct contact interactions  and  $X(3872)$ intermediate state exchanges
in the $s$--channel. However, as we will see later, this does not improve the total $\chi^2$ with respect to Case I.

\end{itemize}

In next subsection
we will try to examine which of the above scenario is favored by  experimental data.

\subsection{Data fitting}

 Using Eqs. (\ref{invariantmassddstar}) and (\ref{invariantmass:Jpipip}),
 we proceed now to fit two sets of  $D^0\bar D^{*0}$ data and two  of $J/\Psi\pi^+\pi^-$
 data. The two sets of $D^0\bar D^{*0}$ data are:
 $B^+\rightarrow XK^+\rightarrow  D^0\bar D^{*0}K^+$
 from BELLE~\cite{BELLE2} and the
$B^+\rightarrow XK^+\rightarrow D^0\bar D^{*0}K^+$
 {}from BABAR~\cite{BABAR4}. The $D^{*0}$ and $\bar D^{*0}$
 are reconstructed from $D^0 \pi^0, D^0\gamma$ and $\bar D^0\pi^0, \bar
 D^0\gamma$, respectively. We perform our fits from
 the $D^0\bar D^{*0}$ threshold up to $\sqrt{s}=3893.8$~MeV for BELLE and $\sqrt{s}=3895$~MeV for BABAR.
 There are also two $J/\Psi\pi^+\pi^-$ data samples (BELLE~\cite{BELLE4} and BABAR~\cite{BABAR3}).
 We fit  from $\sqrt{s}=3843.4$~MeV up to $3892.4$~MeV for BELLE~\cite{BELLE4} and from
 $\sqrt{s}=3847.2$~MeV up to 3897.6~MeV for BABAR~\cite{BABAR3}.

  As explained in the previous subsection, we consider three fit  cases:
    Pure elementary particle (Fit I), where we set  $\lambda_2=0$  (and of course $c_0=0$);
  Pure molecule picture (Fit II), where $g_1,\,  g_2$ and $g_4$ are set to zero;
  and a mixing of the elementary particle
  and molecule state (Fit III), where we have all the parameters except $c_0$ in
  Table~\ref{parameters}, with  $\lambda_2$ real. Unfortunately, the Fit III was found to be unstable: Since too many parameters are involved, no convergent solution is found with positive error matrix. The total $\chi^2$ is not meaningfully improved in comparison to Fit I. Hence we will focus mainly on the first two fits and relegate the discussion in the following.

 The most important  parameters for the X(3872) pole position are the two coupling constants $\lambda_2, g_1$
 and $M_X$. The fitting results are presented in Table~\ref{parameters}.
 The $N_{1i}$(i=1,2,3) and $N_{2i}$(i=1,2) are normalization constants for
 the $D^0\bar D^{*0}$ and $J/\Psi\pi^+\pi^-$ processes, and the $c_{1i}$(i=1,2,3) and $c_{2j}$ ($j=1,2$)
 parameterize the  background contributions  for the $B^+\to K^+D^0\bar D^{*0}$ and $B^+\to K^+ J/\Psi\pi^+\pi^-$ data, respectively.
 Since each spectrum has a different normalization constant $N_{ij}$, in general
 it is not really possible to determine  the $N_{ij}$ and the couplings $g_2$, $g_3$ and $g_5$,
 independently.

\begin{figure}[htbp]
\centering
\subfigure[]{
\label{fig:subfig:a}
\includegraphics[height=1.5in,width=2.0in]{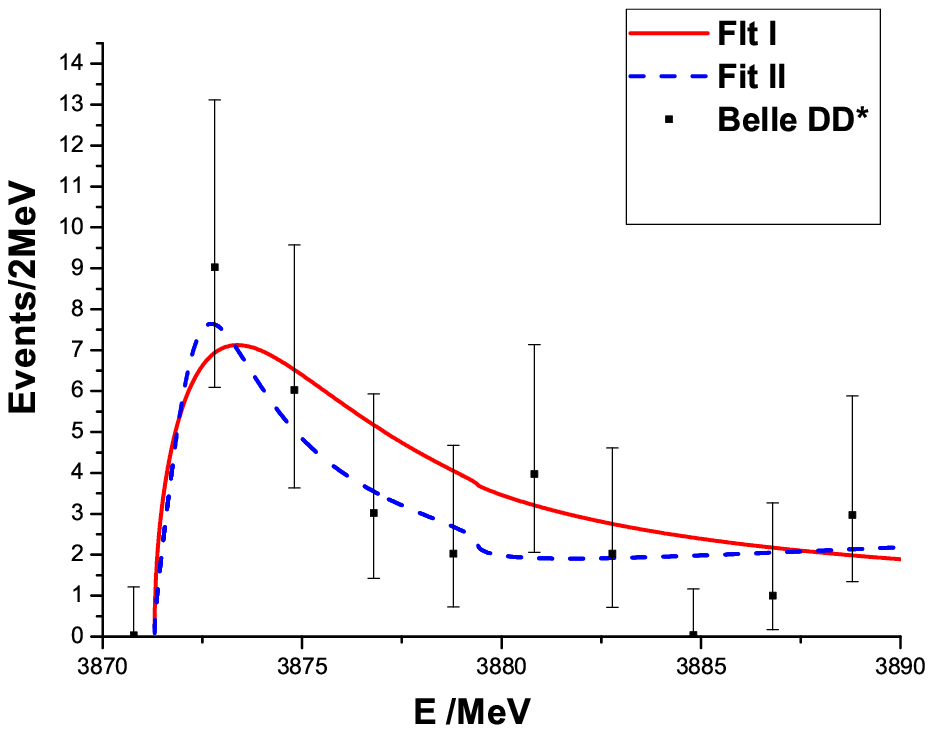}}
\hspace{0.1in}
\subfigure[]{
\label{fig:subfig:b}
\includegraphics[height=1.5in,width=2.0in]{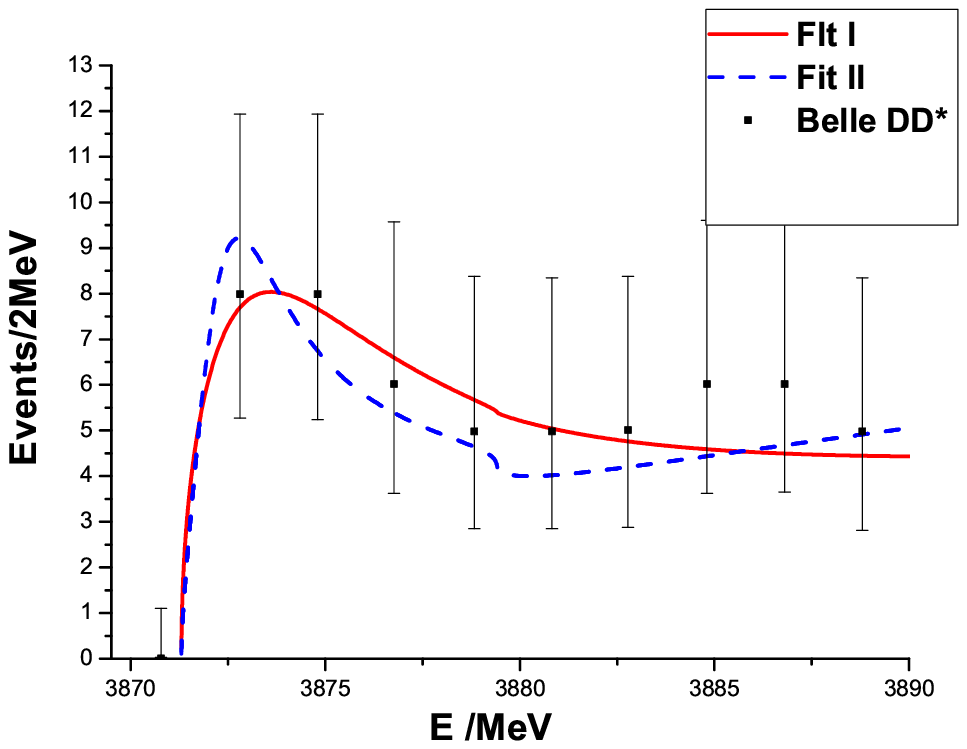}}
\hspace{0.1in}
\subfigure[]{
\label{fig:subfig:c}
\includegraphics[height=1.5in,width=2.0in]{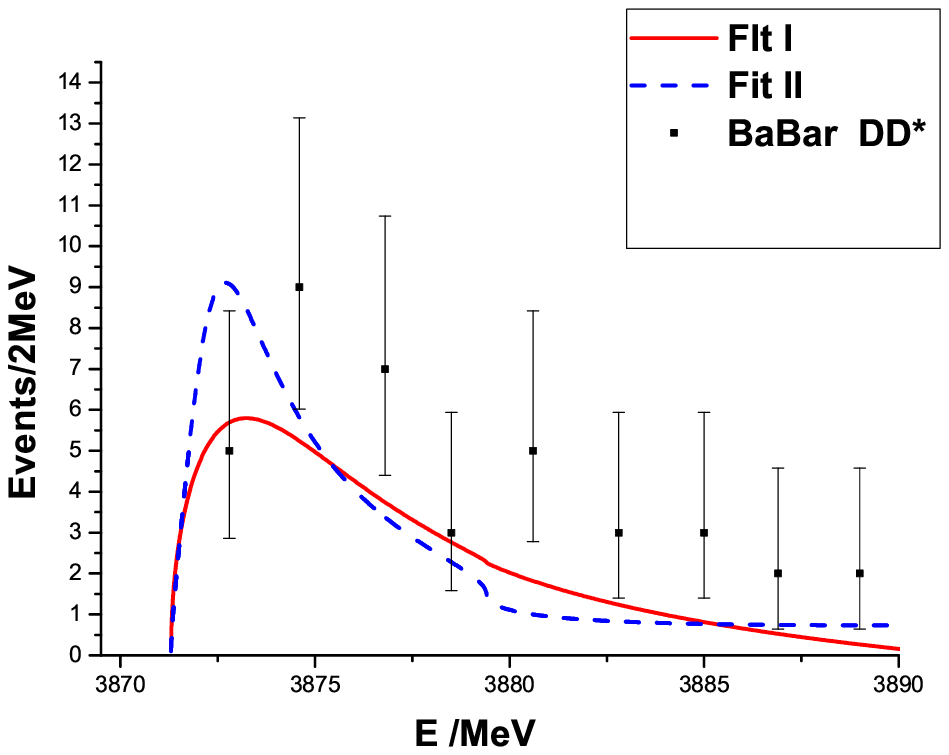}}
\hspace{0.1in}
\subfigure[]{
\label{fig:subfig:d}
\includegraphics[height=1.5in,width=2.0in]{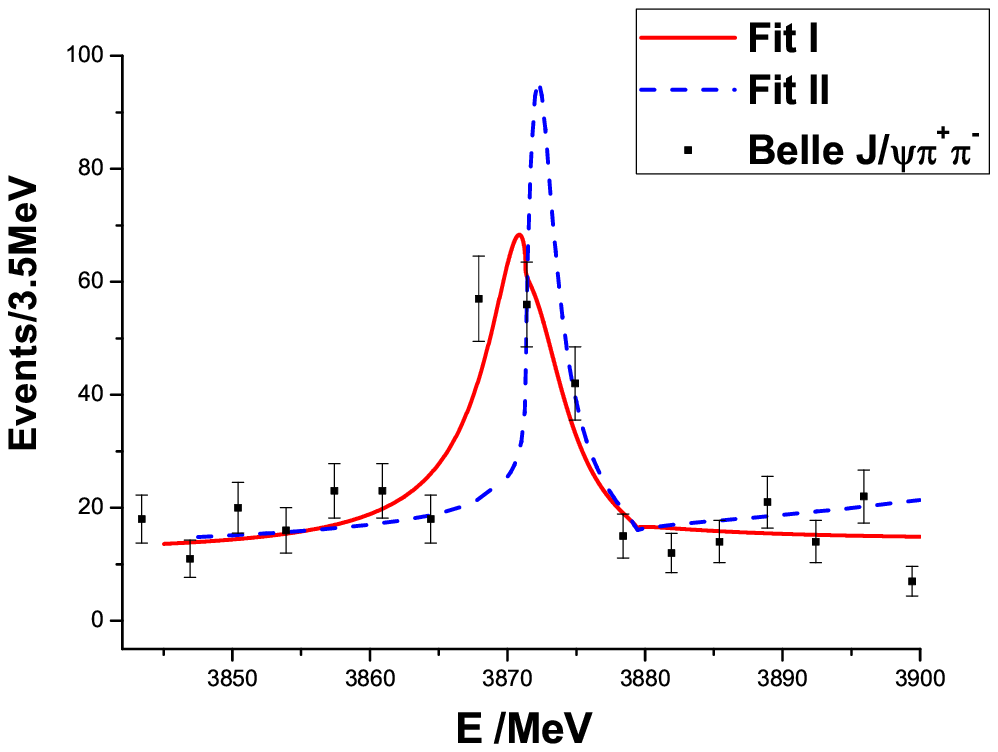}}
\hspace{0.2in}
\subfigure[]{
\label{fig:subfig:e}
\includegraphics[height=1.5in,width=2.0in]{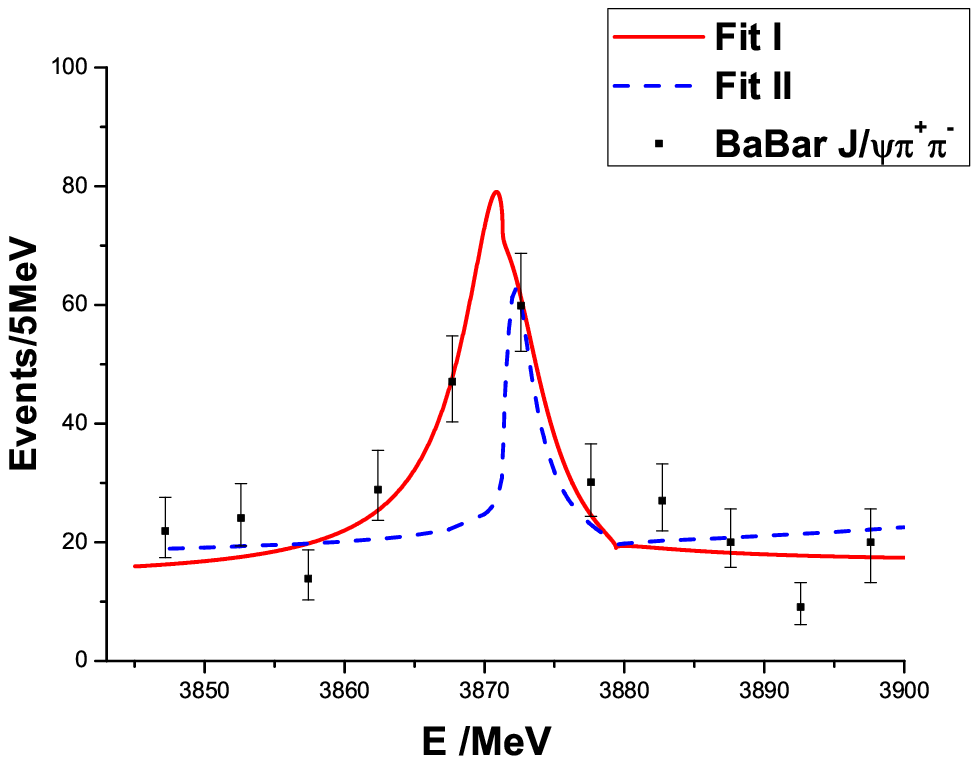}}
\caption[]{(a), (b), $D^0\bar D^{*0}$ invariant mass spectrum for BELLE data. The $\bar D^{*}$ are reconstructed from $\bar D^0\gamma$ and $\bar D^0\pi^0$, respectively in (a) and (b). (c), $D^0\bar D^{*0}$ invariant mass spectrum for BABAR data. (d), (e), $J/\Psi\pi^+\pi^-$ invariant mass spectrum for BELLE and BABAR data, respectively. Here $E=\sqrt{s}$.}
\label{fig:subfig}
\end{figure}

In Table~\ref{parameters}, the $N_{ij} g_3^2$ are obviously larger in Fit II than in Fit I. This is due to the contributions proportional $g_1$ and $g_2$ in Fit I,
which are absent in Fit II and have to be compensated by large values of $N_{1j} g_3^2$
and $N_{2j} g_3^2 g_5^2$.

{\sm{
  \begin{table}[ht]
\begin{center}
 \begin{tabular}  {c c c c}
 \hline\hline
                &     Fit I     &   Fit II      \\ \hline
           &    $\chi^2/dof=47.1/(60-17)$ & $\chi^2/dof=83.3/(60-12)$\\\hline
$\lambda_2$  &      --        &    552.7$\pm$ 1.1      \\
$c_0$         & --&  $ ( 1.70 \pm 0.01) \times 10^{-4}$           \\
$g_1$ (MeV)    &  1977$\pm$908 &   --           \\
$g_2/g_3$ (MeV)  &   196$\pm$52   &  --       \\
$g_4$   &  0.27$\pm$0.08 &   --        \\
$g_4'$   &    0.44$\pm$0.11 &   --        \\
$g_5$ (MeV$^{-1}$)   &     0.016$\pm$0.014    &    1.0 (fixed)    \\
$M_X$(MeV)   &   3870.3$\pm$0.5  & --   \\
$\Gamma_0$(MeV) &          4.3$\pm$1.5        & --           \\
$N_{11}\cdot g_3^2 $($ 10^{-3}\mbox{MeV}^{-3}$)   &  9.2$\pm$5.0
&159$\pm$55      \\
$N_{12}\cdot g_3^2 $($ 10^{-3}\mbox{MeV}^{-3}$)  &  8.1$\pm$4.0
& 181$\pm$53   \\
$N_{13}\cdot g_3^2 $($ 10^{-3}\mbox{MeV}^{-3}$)  & 9.1$\pm$4.7
& 143$\pm$48    \\
$N_{21}\cdot g_3^2 $ ($ 10^{-5}\mbox{MeV}^{-4}$)   &  4.7$\pm1.3$
& $63\pm 35 $ \\
$N_{22}\cdot g_3^2 $ ($ 10^{-5}\mbox{MeV}^{-4}$)   &  3.9$\pm117$
 & $116\pm 33 $  \\
$c_{11}\times 10^{5}$   &    3.4$\pm$1.7   &     3.6$\pm$1.4        \\
$c_{12}\times 10^{5}$   &    1.9$\pm$1.0   &     0.4$\pm$0.2         \\
$c_{13}\times 10^{5}$   &    1.6$\pm$1.2   &    1.1$\pm$1.0          \\
$c_{21}$   &  15.5$\pm$2.1   &    15.1$\pm$2.0        \\
$c_{22}$   &    13.1$\pm$1.5   &  12.6$\pm$1.4          \\
\hline\hline
 \end{tabular}
 \caption{\label{parameters} {\small Fit parameters  for the two different scenarios. In Fit II
 the entry  $N_{2j} g_3^2$ corresponds to $N_{2j}g_3^2 g_5^2$.   The fits only
 depend on $N_{ij} g_3^2$, $g_2/g_3$  and $g_5$ in  Fit I,
 and  $N_{ij} g_3^2 g_5^2$ for Fit II  ($g_2=0$ in this Fit).
 At the practical level, we will fix $g_5=1$ in the Fit II and the
 entry for $N_{2j}  g_3^2$ must be understood as the fit value for $N_{2j}  g_3^2  g_5^2$.
   }}
   \end{center}
\end{table}
}}

The fits of the theoretical expressions~(\ref{invariantmassddstar}) and (\ref{invariantmass:Jpipip})
to data are shown in Fig.~\ref{fig:subfig}.
Fit I  has a much smaller $\chi^2$ per degree of freedom (d.o.f.)
 than that of Fit II.
This  indicates that the model with  bubble chains with contact $D\bar{D}^*$ rescattering alone
is not favored by  experimental data. It is worth mentioning that,
compared with Fit I, the $\chi^2/\mbox{d.o.f.}$  of Fit III is slightly better, but at present stage we are not able to draw a definitive conclusion from
this study.

One can observe an obvious cusp structure at $\sqrt{s}=3879.4$~MeV
 in Fig. \ref{fig:subfig}. This is  due to  the effect of the $D^{\pm}D^{*\mp}$
 threshold in the coupled channel analysis performed in this article, where the interference with  the charged channel is taken into account~\cite{Braaten2009}. 
 
 The energy resolution was also considered in a comparative fit. Nevertheless, the fit results were not sensitive to it.
We also investigated the longitudinal component of the amplitudes. We find that the   poles  in the longitudinal components are spurious and
 are always found very far away from
 the energy region under study. Hence the longitudinal part can only be considered as a part of the background contribution.

\subsection{Pole analysis}

In the most general case,  Fit III,
the denominators of the transverse   and longitudinal parts  can be written
 as
 \bea\label{Tpropagator}
 (s-M_X^2+iM_X(\Gamma_{J\pi\pi}(s)+\Gamma_{J\pi\pi\pi}(s)+\Gamma_0))(1-i\lambda_2\hat{\Pi}_T(s))-ig_1^2\hat{\Pi}_T(s) \, ,
 \eea
  \bea\label{Lpropagator}
  M_X^2(1-i\lambda_2\hat{\Pi}_L(s))-ig_1^2\hat{\Pi}_L(s) \, .
 \eea
We have the same structure for Fit I, but with $\lambda_2$ set to zero.
On the other hand, in the Fit II case, the $X(3872)$ particle propagator is absent and
we have   transverse and longitudinal denominators of the form
 \bea\label{Tpropagator-FitIII}
 \lambda_2^{-1} + i c_0 \,  -\, i\hat{\Pi}_T(s)  \, ,
 \eea
  \bea\label{Lpropagator-FitIII}
   \lambda_2^{-1} + i c_0 \,  -\, i \hat{\Pi}_L(s)  \, .
 \eea
 Near threshold we have $\hat{\Pi}_T=\frac{1}{16\pi}(-\rho(s)+O(\rho^2(s)))$
 and $\hat{\Pi}_L=\frac{1}{16\pi}(\rho^3(s)+O(\rho^4(s)))$
 (a proof can be found in App.~\ref{appendix B}).
 Thus, the transverse and the longitudinal components have different pole locations
 on the complex energy plane.

 If we focus our attention on just the $D^0\bar D^{*0}$ and $D^+D^{*-}$ channels,
 with the threshold at  $\sqrt{s}=3871.3$~MeV and $\sqrt{s}=3879.4$~MeV, respectively,
 the complex Riemann surface is divided into four sheets.
 The $\rho(s)$--sign prescriptions to pass from one Riemann sheet to another
 are provided in Table~\ref{Riemannsheet2}. One should notice that, in addition to the $D^0\bar{D}^{*0}$ and $D^+\bar{D}^{*-}$ channels, there are $J/\Psi\pi\pi$, $J/\Psi\pi\pi\pi$ and other channels represented by $\Gamma_0$, which have lower thresholds. To simplify the discussion, only the near $X(3872)$ resonance channels $D^0\bar{D}^{*0}$ and $D^+\bar{D}^{*-}$ are considered to classify the Riemann sheets.
 The pole positions  of the transverse part
 are presented in Tables~\ref{Tpartpole}. Poles from the longitudinal part are far away from the physical region  and, hence, are spurious and have noting to do with physics under concern.

 \begin{table}[t!]
\begin{center}
 \begin{tabular}  {c c c c c}
 \hline\hline
           &   sheet I    &   sheet II     &   sheet III    & sheet IV  \\ \hline
$\rho_{D^0D^{*0}}(s)$    &  +           &  -             &   -            &  +    \\
$\rho_{D^+D^{*-}}(s)$    &  +           &  +             &   -            & -             \\
\hline\hline
 \end{tabular}
 \caption{\label{Riemannsheet2} {\small
  Definition of the four Riemann sheets with
 the  $D^0\bar D^{*0}$ and $D^+D^{*-}$ channels.
 The $\rho_{D^0D^{*0}}(s)$ and $\rho_{D^+D^{*-}}(s)$
 represent the $\rho(s)$ for each channel in $\hat{\Pi}_T$ and $\hat{\Pi}_L$.  }}
   \end{center}
\end{table}

 \begin{table}[t!]
\begin{center}
 \begin{tabular}  {c c c c}
 \hline\hline
                           Sheet       &   Fit I     & Fit II      \\ \hline
I           & 3871.1-3.3i           &--            \\
 II     &3870.5-3.7i  &3871.7-0.9i      \\
 III     & 3869.0-4.0i  &--     \\
 IV         & 3869.8-3.5i                     &-- \\
\hline\hline
 \end{tabular}
 \caption{\label{Tpartpole} {\small
 Transverse pole position of the $X(3872)$ for the two fits.
 The value of ${  \sqrt{s_{\rm pole}}=M_{\rm pole} - i\Gamma_{\rm pole}/2  }$
 is given in MeV. No pole is found in the sheets I, III and IV for Fit II. Since there are lower thresholds, such as $J/\Psi\gamma$, the pole on sheet I does not break the causality.
 }}
   \end{center}
\end{table}


  In Fit I,  four poles are found, with similar widths of approximately 6 MeV, mainly generated from the elementary component of the X(3872).
 Other channels ($J/\Psi\pi\pi, J/\Psi\gamma$) were also taken into account in Fit I  through
 the $\Gamma_{J/\Psi\pi\pi}$, $\Gamma_{J/\Psi\pi\pi\pi}$ and the parameter $\Gamma_0$ in the propagator of the $X(3872)$, respectively.  They have much lighter thresholds
 than the $D^0\bar D^{*0}$ one and vary smoothly in the small energy region under study.
  A large elementary particle component for the $X(3872)$ is hinted by the poles on the four Riemann sheets that can be found in Table III for Fit I,
 in agreement to the findings in Ref.~\cite{Dailingyun}.

 In Fit II, there is only one transverse pole, determined through Eq.\,(\ref{Tpropagator-FitIII}) and located on the 2nd Riemann sheet, with a width $\Gamma_{\rm pole}=1.8$~MeV slightly smaller than those in Fit I.

 As discussed above, the transverse part of the loop has the near threshold behaviour
 $\hat{\Pi}_T\propto -\rho(s)$. Thus, the coupling  $\lambda_2$
determines  the pole position.
In the Fit II scenario, we only find a pole in the 2nd Riemann sheet, which means the $X(3872)$ would not be a bound state but a virtual state.

\subsection{Fit III Pole Moving}

In the sections above, only Fit I and Fit II were discussed. Instead, Fit III (a combination of Fit I and Fit II) does not give a very different $\chi^2/\mbox{d.o.f.}$ comparing with Fit I, which changes from $47.1/43$ to $42.4/42$. The parameters $g_1=2320\pm 291 \mbox{MeV}$, $g_4=0.18\pm0.09$, $g_4'=0.32\pm0.27$, $M_X=3874.2\pm 0.8 \mbox{MeV}$ and $\Gamma_0=1.7\pm 1.4\mbox{MeV}$ in Fit III are also similar to those in Fit I, but with the additional coupling $\lambda_2=647.1\pm26.0$ in Fit III. However, compare with Fit I, the transverse part of the amplitude has an additional pole on the 1st sheet, which is very close to the real axis. That might provide a different physical picture for the nature of the X(3872) if proven true.

 \begin{figure}[htbp]
\centering
\subfigure[]{
\label{fig:subfig:sheet1}
\includegraphics[height=2.0in,width=3.0in]{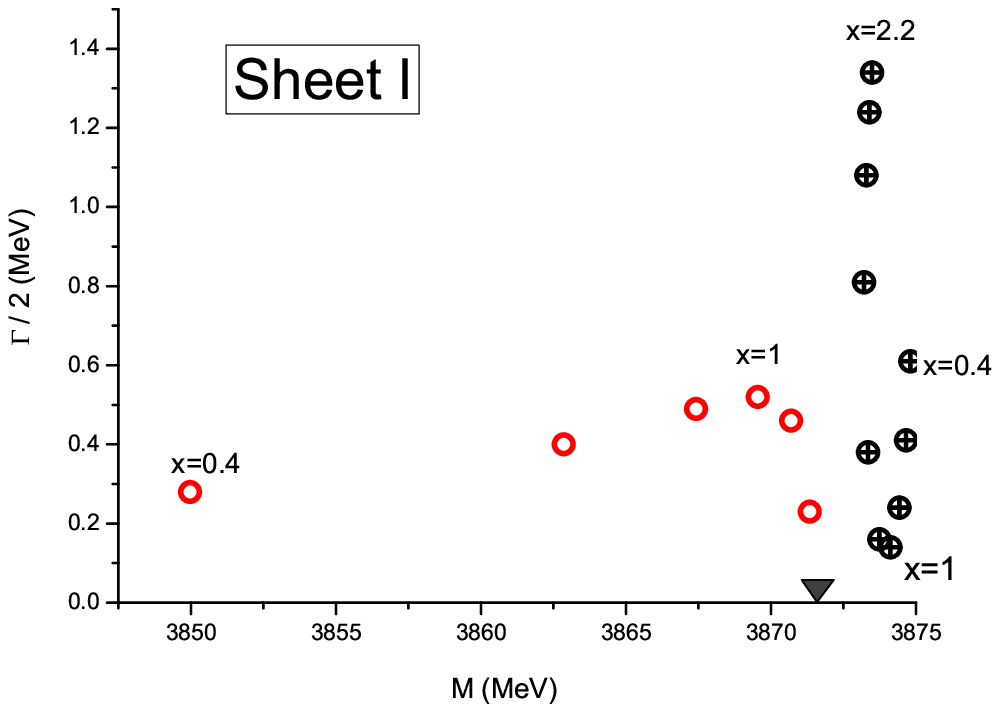}}
\hspace{0.1in}
\subfigure[]{
\label{fig:subfig:sheet2}
\includegraphics[height=2.0in,width=3.0in]{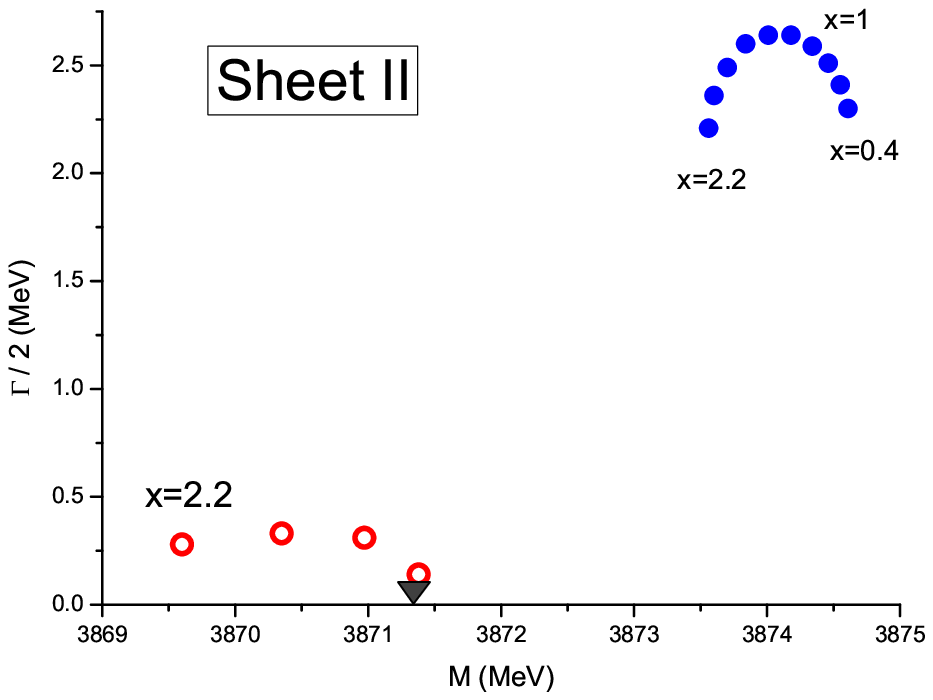}}
\caption[]{Pole trajectories for $x$ varied in the range $[0.4,\,2.2]$, with $\lambda_2=647.1x$. All the other parameters remain fixed at their respective values from Fit III. $x=1$ corresponds to the $\lambda_2$ central value in Fit III.}
\label{polemoving}
\end{figure}

By searching the poles the transverse propagator in Fit III in Eq.\,(\ref{Tpropagator}), we found there is one pole on the first sheet ($\sqrt{s}=3874.1-0.5i$~MeV) very close to the real $s$--axis.   In addition to this pole, there is another pair of poles on the 1st and 2nd sheet at $\sqrt{s}=3869.5-0.1i$~MeV and $3874.3-2.6i$~MeV, respectively, which are also closer than those of Fit I. The other two poles on sheet III and IV have similar positions as in Fit I but slightly smaller imaginary parts, which are around 2.0~\mbox{MeV}. Obviously, the presence of poles on the first sheet would lead to important effects which should be analyzed carefully.

Since the additional pole on the first sheet is the main difference between Fit III and Fit I, one straightforwardly may suppose that the additional pole comes from $D\bar D^*$ rescattering. To confirm this guess, we vary the coupling constant $\lambda_2$ to find out whether the poles are affected by this coupling constant.  We define the scale factor $x$ in the form $\lambda_2=647.1x$ (with $x=1$ corresponding to the central value in Fit III), which is varied in the range $x\in [0,4,\,2.2]$ in steps of $0.2$. The trajectories of poles are shown in Fig. \ref{polemoving}. It reveals that although the elementary $X(3872)$ plays a main role in the propagator, the contact $D\bar D^*$ rescattering becomes important  as $x$ increasing. There is a 1st sheet pole from far below the $D^0\bar D^{*0}$ threshold when $x$ is small and moves to it when $x\approx 1$. If x further increases this 1st sheet pole acrosses the $D^0\bar D^{*0}$ brach cut point in the real axis to the 2nd Riemann sheet and  goes away(the trajectory of this pole is marked with red empty circles in Figs.\,4(a) and 4(b)). It then moves again far away from the region we are studying if we keep increasing x. The other pair of poles on the 1st and 2nd sheets (black cross circle and full blue circle) move just a few MeV when we vary $x$.

    The main effect of bubble chain in Fit III is to bring a new narrow resonance pole below the $D^0\bar D^{*0}$ threshold and cause a sharp spike in the $J/\Psi\pi\pi$ decay spectrum (it could not be seen in $D\bar D^*$ decays). We hope that data with better energy resolution can eventually discern the presence or no of this additional narrow resonance pole.

\section{Conclusion}
In this article we explored  the nature of the $X(3872)$ through an effective Lagrangian model
to describe the energy region around the $D\bar{D}^{*}$ thresholds.
The decays $B^+\to K^+ D^0\bar{D}^{*0}$ and $B^+\to K^+J/\psi \pi^+\pi^-$ were analyzed.
We investigated whether the $X(3872)$ resonance  is mainly an elementary particle, a $D\bar{D}^*$ molecule or
and admixture  of both.
In the analysis where it is assumed to be a pure $c\bar c$ state with  no $D\bar D^*$ component
(Fit I),
the $DD^*DD^*$ four particles coupling constant $\lambda_2$ is set to zero.
The $D\bar{D}^{*}$  rescatterings through an  intermediate $s$--channel $X(3872)$ propagator reproduces  the line-shape of the spectrum  rather well, indicating that the X(3872) is mainly a standard Breit-Wigner resonance.
In the pure molecule analysis (Fit II), the couplings of the elementary $X(3872)$
with other particles were set to zero ($g_1,\, g_2,\, g_4=0$ in Eq.~(\ref{Tpropagator}))  and,
alternatively,  the coupling constant $\lambda_2$ played the crucial  role in the $B$ decays.
In this case, the $D\bar D^*$  contact final state interaction (ruled by $\lambda_2$) determined
the spectrum line-shape.
We studied also the mixed scenario (Fit III),
with the $X(3872)$ containing  both elementary $c\bar c$ and $D\bar D^*$ molecular  components
which  interact with each other in the $B$ decay. However, the $D\bar D^*$ loops from the
direct $ D\bar{D}^{*} \to D\bar{D}^{*}$  contact interaction may have some non-negligible effects which needs further analysis. Meanwhile, Fit III was found to be unstable with the available data and no definitive conclusion could be extracted. Therefore our fits  tend to favor a mostly  elementary  $X(3872)$.

For Fit II (only bubble chains with contact $ D\bar{D}^{*}$ rescattering),
we only find a pole in the 2nd Riemann sheet, corresponding to a virtual state.
In the favored scenario Fit I,
the lighter channels accounted through $\Gamma_{J/\psi \pi\pi}$, $\Gamma_{J/\psi \pi\pi\pi}$ and $\Gamma_0$  play an important role for the $X(3872)$ pole position, being $D\bar D^*$ contributions subdominant.

Our optimal scenario (Fit I) yields the 1st Riemann sheet pole determination (extracted from the fit parameters through a Monte Carlo simulation)
\bea
M_{X(3872)}^{\rm 1st}\,\,\, =\,\,\, ( \, 3871.2 \, \pm \,0.7\, )\,  \mbox{MeV} \, ,
\qquad\quad
\Gamma_{X(3872)}^{\rm 1st}\,\,\, =\,\,\, ( \, 6.5 \, \pm \,1.2\, )\,  \mbox{MeV} \, .
\eea
This narrow state is very near to the $D^0 \bar{D}^{*0}$ threshold ($M_{D^0}+M_{\bar{D}^{*0}} = 3871.3$~MeV).
Going over the $D^0\bar{D}^{0\, *}$ branch cut point one has access to the 2nd Riemann sheet, where we find a pole with position
\begin{equation}
M_{X(3872)}^{\rm 2nd} \,\,\, =\,\,\, ( \,3870.5 \, \pm \,0.2 \, )\,  \mbox{MeV} \, ,
 \qquad \qquad
\Gamma_{X(3872)}^{\rm 2nd} \,\,\, =\,\,\, ( \,7.9\,  \pm  \,1.6\,)\,  \mbox{MeV} \, .
\end{equation}
This pole is again located in the complex plane in the neighborhood of the $D^0\bar D^{*0}$ threshold. The widths of these 1st and 2nd sheet poles are consistent with the broad structure observed in Fig.\,\ref{fig:subfig}, with a width in the spectrum of the order of 5-10~MeV.

\begin{acknowledgments}
The authors thank Prof.~U.-G.~Mei\ss ner for reading the manuscript.  This work is supported in part by National Nature Science Foundations of China under Contract Nos.10925522 and 11021092, and ERDF funds from the European Commission [FPA2010-17747, FPA2013-44773-P,
SEV-2012-0249, CSD2007-00042] and the Comunidad de Madrid
[HEPHACOS    S2009/ESP-1473].

\end{acknowledgments}

\appendix
\section{The supression of longitudinal part near threshold} \label{appendix A}

 In Sec.~\ref{sec.theory}, the longitudinal part was argued  to be a small quantity compared
 to the transverse part.  Here we provide the proof.
 Since for physical on-shell polarization $p_{D^*}\cdot\epsilon_{D^*}=0$,
 we have that  $p\cdot\epsilon_{D^*}=p_{D}\cdot\epsilon_{D^*}$,
 with $p=p_D+p_{D^*}$. In the $D\bar D^*$ rest-frame one finds
 \bea
 p_{D^*}^{\mu}-\frac{p^{\mu}}{2}&=&(\frac{\sqrt{m_{D^*}^2+|\vec{p}_{D^*}|^2}-\sqrt{m_{D}^2+|\vec{p}_{D^*}|^2}}{2},\vec{p}_{D^*})\nn\\
 &=&(\frac{\Delta_m}{2}(1-\frac{|\vec{p}_{D^*}|^2}{m_D
 m_{D^*}}),\vec{p}_{D^*})\nn\\
 &\approx&(\frac{\Delta_m}{2}
 ,\vec{p}_{D^*}),
 \eea
  where $\vec{p}_{D^*}$ is three momentum of $p_{D^*}$ and  $\Delta_m=m_{D^*}-m_D$.
  All the Lorentz components of $(p_{D^*}^{\mu}- p^{\mu}/2)$ are  very small near the
  $D\bar{D}^*$ threshold. One can see in Eq.~(\ref{DDs}) that
  the transverse part is proportional to  $p_K\cdot \epsilon_{D^*}$
  while the longitudinal component  carries a factor
  $p\cdot \epsilon_{D^*} = (p- 2p_{D^*})\epsilon_{D^*}$,
  which is small  near  the $D\bar D^*$ threshold.
  Another reason for the tiny longitudinal part is that there is no pole in the longitudinal part of the propagator
  in the energy region we are studying (close to the $D\bar D^*$  threshold) and
  the corresponding denominator does not enhance the amplitude  as it occurs
  with the transverse component.

\section{The loop integrals}\label{appendix B}
 In Sec.~\ref{sec.theory}, we make use of the Feynman integral
 \bea
 \int\frac{d^Dk}{(2\pi)^D}\frac{g_{\mu\nu}-\frac{k_{\mu}k_{\nu}}{m_{D^{*0}}^2}}{(k^2-m_{D^{*0}}^2+i\epsilon)((p-k)^2-m_{D^0}^2+i\epsilon)}
 =P_{T\mu\nu}\Pi_{T_{D^0\bar D^{*0}}}+P_{L\mu\nu}\Pi_{L_{D^0\bar D^{*0}}},
 \eea
where $\Pi_{T_{D^0\bar D^{*0}}}$ and $\Pi_{L_{D^0\bar D^{*0}}}$ are
 \bea
\Pi_{T_{D^0\bar D^{*0}}}(P^2)&=&\frac{-i}{16\pi^2}(\frac{1}{2}I_0-\frac{p^2}{2m_{D^*}^2}I_2
+\frac{p^2+m_{D^*}^2-m_D^2}{2m_{D^*}^2}I_1
 -\frac{\frac{1}{3}p^2-(m_{D^*}^2+m_D^2)}{4m_{D^*}^2})\nonumber\\
   &&-\frac{iR}{16\pi^2}(1+\frac{p^2}
 {12m_{D^*}^2}-\frac{m_{D^*}^2+m_D^2}{4m_{D^*}^2}),\nonumber\\
 \Pi_{L_{D^0\bar D^{*0}}}(p^2)&=&\Pi_{T_{D^0\bar D^{*0}}}(p^2)+\frac{i}{16\pi^2}\frac{p^2}{m_{D^*}^2}I_2
+\frac{iR}{16\pi^2}\frac{p^2}{3m_{D^*}^2},
 \eea
with the ultraviolet divergence  $R=-\frac{1}{\epsilon}+\gamma_E-ln 4\pi$ and $\epsilon=\frac{4-D}{2}$.

The values  for  the $I_n$  integrals are
 \bea
 I_n=\int_0^1 x^n ln\frac{m_D^2x+m_{D^*}^2(1-x)-p^2x(1-x)}{\mu^2}dx,
 \eea
 \bea
 I_0&=&-B_0(s),\nn\\
 I_1&=&-\frac{1}{2}(1+\frac{m_{D*}^2-m_D^2}{s})B_0(s)+\frac{1}{2s}(A_{D^*}-A_D),\nn\\
I_2&=&-\frac{1}{3}[(1+\frac{m_{D*}^2-m_D^2}{s})^2-\frac{m_{D^*}^2}{s}]B_0(s)+\frac{1}{3s}(A_{D^*}-2A_D)\nn\\
 &&-\frac{1}{18}+\frac{m_{D*}^2-m_D^2}{3s^2}(A_{D^*}-A_D)+\frac{1}{6s}(m_{D^*}^2+m_D^2).
 \eea
 The A and B functions are:
 \bea
 A_{D^*}&=&-m_{D^*}^2(-1+ln\frac{m_{D^*}}{\mu^2}),\nn\\
 A_{D}&=&-m_{D}^2(-1+ln\frac{m_{D}}{\mu^2}),\nn\\
 B_0(s)&=&2-ln\frac{m_D^2}{\mu^2}+\frac{s+m_{D*}^2-m_D^2}{2s}ln\frac{m_D^2}{m_{D^*}^2}+
 \rho(s)ln\frac{\lambda(s)-1}{\lambda(s)+1},
\eea
with  $\rho(s)=\frac{\sqrt{(s-(m_D+m_{D^*})^2)(s-(m_D-m_{D^*})^2)}}{s}$
and $\lambda(s)=\sqrt{\frac{s-(m_D+m_{D^*})^2}{s-(m_D-m_{D^*})^2}}$.
In the 1st Riemann sheet   we have $\rho(s+i\epsilon) =  |\rho(s)|$ over threshold
and $\rho(s)= i |\rho(s)|$ below.

 We renormalize the amplitude through the threshold substraction,
  \bea
 \hat{\Pi}_{T_{D^0\bar D^{*0}}}(s)=\Pi_{T_{D^0\bar D^{*0} }}(s)-\Pi_{T_{D^0\bar D^{*0} }}(s_{th})\, ,
 \nn\\
  \hat{\Pi}_{L_{D^0\bar D^{*0}}}(s)=\Pi_{L_{D^0\bar D^{*0} }}(s)-\Pi_{L_{D^0\bar D^{*0} }}(s_{th})\,,
 \eea
 where $s_{th}=3871.3\mbox{MeV}$ is the threshold of $D^0\bar D^{*0}$ channel.
  Near this threshold the self-energy shows the behaviour
  \bea
  \hat{\Pi}_{T_{D^0\bar D^{*0}}}(s)=\frac{1}{16\pi}(-\rho(s)+O(\rho^2(s)),\nn\\
  \hat{\Pi}_{L_{D^0\bar D^{*0}}}(s)=\frac{1}{16\pi}(\rho^3(s)+O(\rho^4(s)).
 \eea
For the charged channel we use threshold subtraction with $s_{th}=3879.4\mbox{MeV}$.


\begin{thebibliography}{99}
\bibitem{Doublet}Notice the minus sign in the upper component of the doublet, 
as under isospin transformations 
$\left(  u   \atop d \right) \longrightarrow S\,\left( u \atop d \right)$ 
 one has the conjugate doublet
 $\left( -\bar{d} \atop  \bar u \right)$ transforming like $\left( -\bar{d}\atop \bar u\right) \longrightarrow S\,\left(-\bar{d}\atop \bar u\right)$.
 In our D mesons we have the quark content $D^+\sim c\bar{d}$ and $D^0\sim c\bar{u}$, which gives the isodoublet $D$ in the text~\cite{Bratten}. The hermitian conjugate of the isodoublet transforms with $S^{-1}$ acting from the right and can be used to build isospin invariant operators.
\end{thebibliography}

\end{document}